\documentclass[a4paper,11pt]{article}
\usepackage{graphicx}
\usepackage{hyperref}
\usepackage{amssymb}
\title{Yet another insecure group key distribution scheme using secret sharing}
\author{Chris J. Mitchell\\Information Security Group, Royal Holloway, University of London\\
\url{www.chrismitchell.net}}
\date{16th November 2020}

\begin{document}

\maketitle

\section*{Abstract}

A recently proposed group key distribution scheme known as UMKESS, based on
secret sharing, is shown to be insecure.  Not only is it insecure, but it does
not always work, and the rationale for its design is unsound.  UMKESS is the
latest in a long line of flawed group key distribution schemes based on secret
sharing techniques.

\section{Introduction} \label{Intro}  \label{section-introduction}

There is a long and sad history of insecure group (cryptographic) key
establishment schemes based on secret sharing.  As noted by Boyd and Mathuria,
\cite{Boyd03}, the `idea to adapt secret sharing for key broadcasting seems to
have been first proposed by Laih et al.\ \cite{Laih89}', in a paper published
over 30 years ago.  However, the shortcomings of the approach, and of the many
variants that have been proposed since 1989, have been widely discussed for
almost as long, in particular that:
\begin{itemize}
\item as noted by Boyd and Mathuria, \cite{Boyd03}, a `malicious principal
    who obtains one key gains information regarding the shares of other
    principals', and an outside eavesdropper can also gain this information
    if the old group keys  are revealed;
\item again as noted by Boyd and Mathuria, \cite{Boyd03}, since `knowledge
    of any of the shared secrets is sufficient to construct the session
    key, none of these protocols provides forward secrecy';
\item insider attacks of various attacks appear impossible to prevent, as
    many authors have observed (see, for example,
    \cite{Liu17,Mitchell18,Mitchell19}, and the papers cited therein).
\end{itemize}

The history of such protocols is long and tangled, but one sequence of flawed
protocol proposals, breaks, proposed fixes, and breaks of the fixes is
explained very carefully in Section 5 of Liu et al.\ \cite{Liu17}, and we now
briefly summarise part of the story. In 2010, Harn and Lin \cite{Harn10}
proposed an `authenticated group key transfer protocol based on secret sharing'
(itself intended to address issues in the Laih et al.\ scheme \cite{Laih89}
from 1989). Unfortunately, this was shown not only to be insecure (by Nam et
al.\ \cite{Nam11,Nam12}) but also erroneous in that it does not always work
even if all parties execute it correctly (see Nam et al.\ \cite{Nam12}). Nam et
al.\ \cite{Nam12} also proposed a fixed version, but this was shown to be
insecure by Liu et al.\ \cite{Liu17}.  Inspired by the Harn and Lin 2010 paper,
Sun et al.\ \cite{Sun12} proposed another group key transfer protocol using
secret sharing, and this was shown to be insecure by both Kim et al.\
\cite{Kim13a} and Olimid \cite{Olimid13}.  Olimid \cite{Olimid13} also proposed
a fix, but this was shown to be insecure by Kim et al.\ \cite{Kim13b}.  These
are not the only examples of broken schemes of this type --- one common element
is the lack is a rigorous proof of security in a complexity-theoretic setting,
the established state of the art for such protocols for the last decade or two.

Unfortunately, despite the extensive literature pointing out these and other problems, new and
fundamentally flawed schemes of this general type keep being published. One common element in the
papers published over the last 31 years is that many share one of the authors of the 1989 paper. A
further common element is that each new paper cites some of the previously published schemes, but
many completely fail to acknowledge any of the many attacks against the previously published and
often very closely related schemes. This is most unfortunate, especially given that many of the
newer schemes suffer from the same problems as older schemes.  As we show below, some of the above
statements are also true for UMKESS, a scheme of this general type published in a very recent paper
by Hsu, Harn and Zeng \cite{Hsu20}.

The remainder of the paper is structured as follows.  The UMKESS scheme is
summarised in \S\ref{section-scheme}.  A detailed critique is provided in
\S\ref{section-critique}.  A brief discussion of why proposing arbitrary fixes
to such schemes is unwise is given in \S\ref{section-fixes}, and conclusions
are drawn in \S\ref{section-conclusion}.

\section{The UMKESS scheme}  \label{section-scheme}

\subsection{Objectives}

This scheme is designed to allow a single trusted authority, the \emph{Key
Generation Centre (KGC)} to simultaneously distribute a number of secret group
keys to a number of distinct sets (\emph{groups}) of entities, with each set
being drawn from a larger set of entities all of which have a pre-established
relationship with the KGC.

The scheme uses the Shamir secret sharing scheme \cite{Shamir79}, involving
polynomials over a prime finite field $\mbox{GF}(p)=\mathbb{Z}_p$, for large
$p$.

\subsection{Preliminaries}

Prior to use a large \emph{safe} prime $p$ is selected.  The definition of safe
is not provided by the authors, but presumably it must be sufficiently large to
prevent exhaustive searching for individual keys (which are elements of
GF($p$)).

The protocol involves the KGC and a set of $n$ users
$\mathcal{U}=\{U_1,U_2,\ldots,U_n\}$, from which groups are created who are
provided with new shared session keys by the KGC on demand.  Each user
$U_i\in\mathcal{U}$ is assumed to share a unique secret $x_i\in\mbox{GF}(p)$
with the KGC.

All involved parties must also agree on a cryptographic hash-function $h$,
whose domain and range is GF($p$).

\subsection{Security claims}

The authors claim the protocol is secure against both insider and outsider
attacks, where an insider attacker is a member of $\mathcal{U}$.  The security
properties are not defined formally.

\subsection{Operation}  \label{subsection-operation}

As noted above, the protocol enables the KGC to simultaneously broadcast a set of group keys to a
disparate collection of groups.  We suppose that an instance of the protocol is being executed to
distribute $m$ group keys $K_1,K_2,\ldots,K_m$ to $m$ distinct groups $G_1,G_2,\ldots,G_m$, where
$G_i\subseteq\mathcal{U}$ and we write $|G_i=s_i|$ for every $i$, $1\leq i\leq m$. For each group
$G_i=\{U_{i_1},U_{i_2},\ldots,U_{i_{s_i}}\}$, say ($1\leq i\leq m$), define
\[S(G_i)=\sum_{j=1}^{s_i}i_j \]
i.e.\ $S(G_i)$ is the sum of the indices of the members of the group.  Here as
throughout addition is computed in GF($p$), i.e.\ modulo $p$.

%However, to simplify the
%presentation, we describe its operation in the special case where only two
%groups $G = \{U_{w_1}, U_{w_2}, \ldots, U_{w_s}\},G' = \{U_{w'_1}, U_{w'_2},
%\ldots, U_{w'_{s'}}\} \subseteq \{U_1,U_2,\ldots,U_n\}$ are involved (where
%$1<s,s'\leq n$).  Note that we also need the values $S=\sum_{i=1}^sw_s$ and
%$S'=\sum_{i=1}^{s'}w_{s'}$, which are assumed to be known by the respective
%group members.

The protocol proceeds as follows, where the step numbers correspond to those
given by Hsu et al.\ \cite{Hsu20}.

\begin{enumerate}
\item[2.] The KGC broadcasts the list of groups $G_1,G_2,\ldots,G_m$ and
    their members in a reliable way, i.e.\ it is assumed that these cannot
    be modified by a malicious insider or outsider\footnote{This
    integrity/authenticity assumption is implied but never explicitly made,
    but without it certain obvious attacks apply, as discussed in
    \S\ref{subsection-reliable} below.}.
\item[3.] Each participating user $U_i\in\mathcal{U}$ ($1\leq i\leq n$), i.e.\ each user who is
    a member of at least one group, proceeds as follows. Suppose $U_i$ is a member of $m_i$
    groups $G_{i_1},G_{i_2},\ldots,G_{i_{m_i}}$. $U_i$ chooses $m_i$ random values
    $r_{i_j}\in\mbox{GF}(p)$, $1\leq j\leq m_i$, and sends them (unprotected) to the KGC, i.e.\
    in a way that might permit them to be changed by a malicious party (this assumption is in
    line with the protocol specification --- see, for example, the `proof' of Theorem 5
    \cite{Hsu20}).
\item[4.] Once the KGC has received the sets of random values $r_{i_j}$
    from all the participating members of $\mathcal{U}$, it performs the
    following steps.
    \begin{enumerate}
    \item The KGC chooses $m$ random keys $K_i\in\mbox{GF}(p)$, $1\leq
        i\leq m$, where $K_i$ is intended for use by group $G_i$, and a
        random value $r_0\in\mbox{GF}(p)$.
    \item For each participating user $U_i$ ($1\leq i\leq n$), the KGC:
        \begin{itemize}
        \item computes the unique degree $m_i$ polynomial $f_i$
            over GF($p$) that passes through the following $m_i+1$
            points:
        \[ (i,x_i+r_0)~\mbox{and}~(S(G_{i_j}),K_{i_j}+h(x_i+r_{i_j}+r_0)),~1\leq j\leq m_i; \]
        \item randomly chooses a set of $m_i$ points
            $\{P_1,P_2,\ldots,P_{m_i}\}$ lying on the curve defined
            by $f_i$; and
        \item sends $P_1,P_2,\ldots,P_{m_i}$ to $U_i$ (again
            unprotected, i.e.\ in a way that might permit them to
            be changed by a malicious party).
        \end{itemize}
    \item The KGC makes the values of $r_0$ and $h(K_i)$, $1\leq i\leq
        m$, publicly available to all members of $\mathcal{U}$ in a
        reliable way, i.e.\ it is assumed that these cannot be modified
        by a malicious insider or outsider\footnote{Again this
        assumption is only implicit, but without it certain attacks
        apply --- see \S\ref{subsection-reliable}.}.
    \end{enumerate}
\item[5.] Each participating user $U_i$ ($1\leq i\leq n$) proceeds as follows.
    \begin{enumerate}
    \item On receipt of $P_1,P_2,\ldots,P_{m_i}$, $U_i$ uses them
        together with the point $(i,x_i+r_0)$ to recover the degree
        $m_i$ polynomial $f_i$.
    \item Using $f_i$ and $S(G_{i_j})$, $1\leq j\leq m_i$, $U_i$ can
        compute $K_{i_j}+h(x_i+r_{i_j}+r_0)$ and hence $K_{i_j}$, for
        every $j$.
    \item Finally, $U_i$ checks the recovered group keys $K_{i_j}$
        against the published list of values $h(K_i)$, $\leq i\leq m$,
        made available in a reliable way to all participants.
    \end{enumerate}
\end{enumerate}

In essence, a separate `secret' polynomial is computed for each participating
user, and the user recovers group keys from points on this polynomial (which
has degree equal to the number of group keys to be distributed to this user).

\section{A critique}  \label{section-critique}

\subsection{A definitional issue}

We first observe that, in certain not unlikely cases, the system cannot work.

In Step 4(b), the KGC generates the following $m$ points:
\[ (S(G_{i_j}),K_{i_j}+h(x_i+r_{i_j}+r_0)),~1\leq j\leq m_i; \]
Clearly, if the values $r_{i_j}$ are all distinct, $1\leq j\leq m_i$, then the
$y$ coordinates will all be distinct.  However, there is nothing to prevent the
possibility that $S(G_{i_j})=S(G_{i_{j'}})$ for two distinct groups $G_{i_j}$
and $G_{i_{j'}}$.  This could happen very easily, e.g.\ if
$G_{i_j}=\{U_1,U_5\}$ and $G_{i_{j'}}=\{U_1,U_2,U_3\}$, where we have
$S(G_{i_j})=S(G_{i_{j'}})=6$. In such a case, the polynomial $f_1$ for user
$U_1$ cannot exist, since it cannot pass through two points with the same $x$
coordinate but distinct $y$ coordinates.

This issue could, of course, be fixed, e.g.\ by replacing $S(G_i)$ throughout
by a unique numeric identifier for the group $G_i$.  Indeed, it would seem
reasonable to require the KGC to devise a new (and unique) set of group
identifiers for every instance of the protocol, and to distribute them as part
of Step 2 of the protocol.  However, given that there are more serious issues
with the security of, and rationale for, the protocol, we do not explore such
fixes further here.

\subsection{A serious security weakness}  \label{subsection-attack}

We now demonstrate that a much more serious security issue exists, in that the
long-term secret $x_i$ of one user can be recovered by another user (an insider
attacker), who needs only make a small modification to one message sent to the
KGC by the `victim' user and then intercept the response.  We use the same
notation as employed in the protocol description in
\S\ref{subsection-operation}.

We suppose that the insider attacker ($U_a$, say) is a member of (at least) two
groups in common with the victim user $U_v$.  Suppose that $U_a$ intercepts the
set of random values $\{r_1,r_2,\ldots,r_{m_v}\}$ sent by user $U_v$ to the KGC
in Step 3, and prevents them reaching the KGC; we suppose also, without loss of
generality, that $U_a$ is a member of the two groups $G_{v_1}$ and $G_{v_2}$.
We further suppose that $T$ modifies the set of random values sent by $U_v$ to
$\{r_1,r'_2,r_3,\ldots,r_{m_v}\}$ before forwarding them to the KGC, where
$r'_2=r_1$.

The protocol proceeds exactly as specified and we observe that $U_a$, as a
legitimate protocol participant, will be able to learn $K_{v_1}$ and $K_{v_2}$
from the set of points it is sent by the KGC (since we assumed that $U_a$ is a
member of the two groups $G_{v_1}$ and $G_{v_2}$).

We further suppose that $K_a$ intercepts the set of points
$P_1,P_2,\ldots,P_{m_v}$ sent to $U_v$ --- these points will all lie on the
polynomial $f_v$ generated by the KGC in Step 4.  This polynomial will also
pass through the points:
\[ (S(G_{v_1}),K_{v_1}+H)~\mbox{and}~(S(G_{v_2}),K_{v_2}+H) \]
(amongst others), where $H=h(x_v+r_{v_1}+r_0)$.  That is, apart from the $m_i$
points $P_1,P_2,\ldots,P_{m_v}$, $U_a$ will know the difference between the $y$
values for two other points on the curve defined by $f_v$ (with known $x$
values).  That is, if we let $z_1=S(G_{v_1})$ and $z_2=S(G_{v_2})$, $U_a$ will
know the following equation holds:
\[ f_v(z_1)-f_v(z_2) = K_{v_1}-K_{v_2} \]
where all the values (apart from the coefficients of $f_v$ are known. This
yields a linear equation in the coefficients of $f_v$.

The $m_v$ points $P_1,P_2,\ldots,P_{m_v}$ can be used to yield a set of $m_v$
further linear equations in the $m_i+1$ coefficients of $f_v$, i.e.\ $U_a$ will
have a set of $m_v+1$ linear equations in the $m_v+1$ coefficients of $f_v$,
which will almost certainly be independent given that $P_1,P_2,\ldots,P_{m_v}$
are randomly chosen and $p$ is very large.  These can very easily be solved to
yield $f_v$.  Finally, $U_a$ simply evaluates $f_v(v)$ to yield $x_v+r_0$,
i.e.\ $U_a$ has the long-term secret of $U_v$ (since $r_0$ is public).

That is, using this simple attack, one legitimate user can obtain the secret
belonging to another user, and can thereafter learn all the group keys issued
to this user.  This clearly invalidates Theorem 5 of Hsu et al.\ \cite{Hsu20};
this is not so surprising since the `proof' offered is a series of heuristic
arguments rather than a rigorous proof.

\subsection{Reliable broadcasts}  \label{subsection-reliable}

In the protocol description in \S\ref{section-scheme}, there are four main
communications flows:
\begin{itemize}
\item two broadcasts to all participants from the KGC: a broadcast of the
    list of groups (Step 2), and a broadcast of the values $r_0$ and
    $h(K_i)$ ($1\leq i\leq m$) (Step 4c);
\item transmission of $m_i$ random values $r_{i_j}$ from each participating
    user $U_i$ to the KGC (Step 3);
\item for every participating user $U_i$, transmission from the KGC to
    $U_i$ of the set of points $\{P_1,P_2,\ldots,P_{m_i}\}$ (Step 4b).
\end{itemize}
Hsu et al.\ \cite{Hsu20} do not made clear the degree to which these
communications flows need to be protected.  They variously refer to a
`broadcast channel', `broadcasts', and making information `publicly known'.
However they do claim (in the `proof' of Theorem 5), that 'service requests
from group members are not authenticated', and also that `an adversary
(insider) can $\ldots$ forge challenges of other group member'.  They also
explicitly refer to the possibility that one of the $m_i$ values $r_{i_j}$ is
modified by an adversary.

We have therefore assumed throughout this paper that the transmission of the
$r_{i_j}$ values to the KGC in Step 3 is unprotected.  This enables the attack
described in \S\ref{subsection-attack}.  We have correspondingly assumed that
the transmission of the points $P_1,P_2,\ldots,P_{m_i}$ from the KGC to each
participating user $U_i$ in Step 4b is unprotected, although we do not discuss
this further here.

There is no substantive discussion of the security requirements for the two
broadcasts made by the KGC to all participants.  On reflection, and to be as
fair as possible to the protocol designers, we have assumed that these are
protected in some way, e.g.\ by being posted on a KGC website which can be
authenticated (e.g.\ using TLS).  Of course, this adds an `invisible' overhead
to the protocol, but it is a necessary assumption, since if either of these
broadcasts can be manipulated then attacks are possible, as we now briefly
describe.

\begin{itemize}
\item If the list of groups can be manipulated then a simple outsider
    attack is possible which we describe in the form of a short example.
    Suppose group $G_i$ in the list includes the users $U_1$, $U_2$ and
    $U_3$. Then, clearly, $S(G_i)=6$. Suppose that the version of the group
    list sent to $U_1$ is modified to $G'_i$ so that $G'_i$ includes $U_1$
    and $U_5$.  Then $S(G'_i)=6$, i.e.\ the polynomial $f_1$ computed by
    the KGC would be exactly the same in both cases; this means that, when
    performing the protocol, user $U_1$ will recover key $K_i$ correctly,
    but will believe it is shared with user $U_5$ when it is in fact shared
    with users $U_2$ and $U_3$.  This is clearly not a desirable situation.

\item If the list of hashed keys $h(K_i)$ ($1\leq i\leq m$) can be
    manipulated, then in this case an insider attack is possible, which we
    again describe in the form of a simple example.  Suppose a `victim'
    user $U_v$ is in the same group, $G_{v_t}$ say (for some $t$ satisfying
    $1\leq t\leq m_v$), as an attacker user $U_a$. Both users perform the
    protocol correctly, except $U_a$ prevents the correct list of hashed
    keys $\{h(K_1), h(K_2), \ldots, h(K_m)\}$ and the correct set of points
    $\{P_1,P_2,...,P_{m_v}\}$ reaching $U_v$.  $U_A$ completes the protocol
    correctly, and learns $K_{v_t}$ (since both $U_a$ and $U_v$ are in
    group $G_{v_t}$). $U_a$ now chooses a key $K'_{v_t}$ which will be
    accepted by $U_v$ instead of $K_{v_t}$.

$U_a$ next computes the unique polynomial $\delta$ of degree $m_v$ passing
    through the $m_v+1$ points:
\[ (i,0),~(S(G_{v_t}),K'_{v_t}-K_{v_t})~\mbox{and}~(S(G_{v_j}),0),~1\leq j\leq m_i~(j\not=t). \]
Suppose the points $P_1,P_2,...,P_{m_v}$ sent by the KGC to $U_v$ (but
which did not reach $U_v$) satisfy $P_i=(x_i,y_i)$. $U_a$ now computes a
new set of points $D_i=(x_i,d_i)$, $1\leq i\leq m_v$, which lie on
$\delta$, and puts $P'_i=(x_i,y_i+d_i)$, $1\leq i\leq m_v$.  It should be
clear that the points $P'_i$ all lie on the curve defined by the polynomial
$f_v+\delta$; it should also be clear that the point $(v,x_v+r_0)$ also
lies on this curve, although the $y$ value is of course not known to $U_a$.

$U_a$ now sends to $U_v$ (masquerading as the KGC), the correct set of
hashed keys except that $h(K_{v_i})$ is replaced by $h(K'_{v_i})$, and the
new set of points $P'_1,P'_2,...,P'_{m_v}$.  Since $P'_1,P'_2,...,P'_{m_v}$
and $(v,x_v+r_0)$ all lie on the curve defined by $f_v+\delta$, this is the
polynomial that will be recovered by $U_v$ (instead of $f_v$).  $U_v$ now
evaluates this polynomial and it is simple to see that $U_v$ will recover
the correct set of keys except that $K_{v_i}$ will be replaced by
$K'_{v_i}$ --- this is consistent with the manipulated set of hashed group
keys, and hence $U_v$ will accept the recovered keys as valid.
\end{itemize}

\subsection{A questionable rationale}

We further point out that the rationale for the scheme is highly questionable.
One instance of the scheme costs each participant a total of $m_i$ executions
of the hash function $h$, together with solving for the coefficients of a
degree $m_i+1$ polynomial and a few modular additions, i.e.\ on average one
hash execution plus some minor computations for each key.

Hsu et al.\ \cite{Hsu20} compare the cost of their scheme with two other
protocols. The first uses a public key cryptosystem, and the second involves a
number of parallel executions of another secret sharing based scheme proposed
by Harn and Lin \cite{Harn10}. Neither of these are sensible comparisons.  The
public key scheme is designed with different assumptions, and it would be
expected to be significantly more costly.  The comparison with the scheme of
Harn and Lin makes no sense at all because, as discussed in
\S\ref{section-introduction}, it is known to be insecure.  Moreover, the
comparisons avoid the cost of providing publicly verifiable lists of the groups
and of group key hashes.

Even more importantly, there are very well-established protocols which achieve
the same goal in a provably secure way at comparable computational and
communications cost, and which avoid the need for a publicly verifiable
publication of group key hashes.  The authors completely ignore the huge and
very well-established literature in the area, e.g.\ as summarised in the
excellent Boyd and Mathuria \cite{Boyd03} (and the recent second edition,
\cite{Boyd20}). Indeed, there is even an international standard for group key
establishment --- ISO/IEC 11770-5 \cite{ISO11770-5:11}, which was published in
2011.

\section{Pointless fixes}  \label{section-fixes}

In \S\ref{section-introduction}, some of the sad history of group key
distribution schemes based on secret sharing was described.  It seems clear
that the cycle of design, break and fix is itself broken, at least until and
unless a `fixed' protocol is proven secure in a rigorous way.  This point is
made by Liu et al.\ \cite{Liu17}.

\begin{quote}
The security proof for each vulnerable group key distribution protocol only
relies on incomplete or informal arguments.  It can be expected that they would
suffer from attacks.
\end{quote}

Sadly, this lesson has not yet been recognised by everyone.  Apart from the
cases mentioned in \S\ref{section-introduction}, we should also mention the
secret-sharing-based group key transfer scheme proposed by Hsu et al.\ in 2017
\cite{Hsu17}.  This was shown to be insecure \cite{Mitchell18} shortly after
its publication.  In a response published shortly afterwards, Kisty and Saputra
\cite{Kisty18} proposed a fixed version of the 2017 protocol.  Sadly this `fix'
completely lacks a rigorous security analysis.  As a result, it too may be
insecure.  However, perhaps more significantly, the fix involves the addition
of digital signatures to enable recipients of certain messages to verify their
origin and integrity.  Whilst this may well prevent attacks, it completely
negates any rationale for the design of the protocol by greatly increasing the
computational complexity.  Distributing group keys using public key techniques
is a well known and solved problem, and thus the Kisty-Saputra scheme is not a
valuable contribution to the literature.

Finally we observe that there are well-established formal security models within which properties
of group key establishment protocols can be established --- see in particular Bresson et al.\
\cite{Bresson02} and Gorantla et al.\ \cite{Gorantla11}. A helpful summary of the scope of the
various models for group key establishment protocols can be found in \S 2.7.1 of Boyd et al.\
\cite{Boyd20}.

\section{Concluding remarks}  \label{section-conclusion}

In this paper we have discussed two related themes: the (sad) history of
insecure group key distribution schemes based on secret sharing, and the
details of why a specific example of a recently proposed scheme of this type is
insecure. Perhaps the saddest point is that the literature reviewed here is
only a small sample of a very extensive literature on secret-sharing-based
group key distribution, including a number of other sagas involving schemes
repeatedly broken and fixed.

In conclusion, this evidence strongly argues in favour of two recommendations.
Firstly, the academic world should stop publishing security schemes for which
there is a lack of robust evidence of security.  Secondly, academia should stop
attempting to publish fixed schemes which are pointless either because there is
no proof of security or because, whilst they may be secure, they invalidate the
rationale of the original unfixed scheme.

%\bibliographystyle{plain}
%\bibliography{Crypto}

\end{document}